\documentclass[%
 aip,
 pop,
 amsmath,amssymb,
 reprint
]{revtex4-1}

\usepackage{graphicx}
\usepackage{dcolumn}

\usepackage{bm}

\usepackage[utf8]{inputenc}
\usepackage[T1]{fontenc}
\usepackage{mathptmx}
\usepackage{color}

\newcommand{\pderv}[2]{\frac{\partial #1}{\partial #2}}

\newcommand{\rem}[1]{}

\begin{document}

\title{Zebra-like patterns in {whistler wave} emission spectra from nonequilibrium mirror-confined laboratory plasma}

\author{M. E. Viktorov}
\email[Author to whom correspondence should be addressed. Electronic mail: ]{mikhail.viktorov@appl.sci-nnov.ru}
\author{A. G. Shalashov}
\author{E. D. Gospodchikov}
\affiliation{Institute of Applied Physics of the Russian Academy of Sciences, 46 Ulyanov str., 603950 Nizhny Novgorod, Russia}
\author{N. Yu. Semin} 
\affiliation{Institute of Applied Physics of the Russian Academy of Sciences, 46 Ulyanov str., 603950 Nizhny Novgorod, Russia}
\affiliation{Lobachevsky State University, 23 Gagarina av., 603950 Nizhny Novgorod, Russia}
\author{S. V. Golubev}
\affiliation{Institute of Applied Physics of the Russian Academy of Sciences, 46 Ulyanov str., 603950 Nizhny Novgorod, Russia}

\date{\today}

\begin{abstract}
Zebra-like patterns have been observed in the electron cyclotron emission spectra from strongly nonequilibrium plasma confined in a table-top mirror magnetic trap. The analysis of the experimental data suggests that the formation of zebra-like patterns could eventually be related to modulation of the whistler waves by the  {ion-acoustic} waves  excited during the abrupt ejection of electrons into a loss cone caused by the development of the whistler  instability under the electron cyclotron resonance condition. 
\end{abstract}


\maketitle 

\section{Introduction}

The dynamics of energetic charged particles in space and laboratory magnetic traps defines the spectra of electromagnetic waves excited as a result of different types of kinetic instabilities. 
In some cases, the spectrum of excited waves is very wide and contains regular variations, such as frequency sweeping events. In a magnetoactive plasma a certain set of harmonics in a dynamic spectrum of plasma electromagnetic emission is often observed, which could be naturally explained as electron or ion cyclotron instability development. In space plasmas, the excitation of waves at cyclotron harmonics is observed in numerous systems, such as magnetosphere of the Earth \cite{Ronnmark_1978, Mazouz_2009} and other planets \cite{Tao_GJR_2010}. Under laboratory conditions, the excitation of cyclotron harmonics by fast particles is observed both in linear \cite{Ikezoe_2015, Compernolle_PRL_2015, Van_Compernolle_2017} and toroidal \cite{Sharapov_2013} magnetic traps. 

Frequently, the observation of banded harmonics in plasma electromagnetic emission is not directly related to a cyclotron instability but explained by the excitation of plasma waves under the condition of double plasma resonance which is very typical for astrophysical plasmas \cite{Zheleznyakov_UFN_2016}. This phenomenon was firstly observed in the Sun radio emission \cite{Zheleznyakov_1975a,Zheleznyakov_1975b}, but also presents in the Jovian decameter radio emission \cite{Kurth_2001,Litvinenko_2016}, in the VLF hisses of the terrestrial magnetosphere \cite{Titova_GRL_2007} and in a laboratory mirror-confined plasma \cite{Mansfeld_DPR_2018}.
 
In all the above cases, the spectrum of plasma emission forms zebra-like patterns in a time-frequency space. Similar kinds of spectral features could be also related to the variations of the medium through which the electromagnetic wave propagates.  
It is known that the continuous variation of plasma electron density would cause modulations to both amplitude and phase of the transmitter electromagnetic wave \cite{Yang_2015}. In this case, the spectrum of transmitted monochromatic wave will be enriched with harmonics at the modulation frequency and the modulation strength is determined by the ratio between the plasma frequency and the incident wave frequency.
{Modification of whistler wave spectrum due to propagation in a plasma with spatially modulated parameters is well studied in laboratory experiments \cite{Kostrov_2003,Aidakina_2015} and ionosphere heating experiments \cite{Gurevich_UFN_2007,Golkowski_GRL_2009,Sharma_JGR_2010}}.

In the present paper, we study the electromagnetic emission spectrum of plasma which is created and sustained by high-power microwave radiation of a gyrotron under the electron cyclotron resonance (ECR) condition in a simple mirror trap \cite{Viktorov_EPL_2015}. The detailed overview of possible types of kinetic instabilities in the present experiment has been given in Ref.~\onlinecite{Shalashov_2017_PhP}. The focus of this work is given to the study of time-frequency characteristics of plasma microwave emission 
in the whistler wave frequency domain, i.e. below electron cyclotron frequency. The distinguishing feature of this kind of emission is a zebra-like dynamic spectrum with a few narrow-band stripes which are placed equidistant relative to each other. We will study the emission spectrum and discuss shortly some possible reasons of zebra-like patterns. 

\section{Experimental setup and diagnostics}

\begin{figure*}[tbp]
	\centering
	\includegraphics[width=160mm]{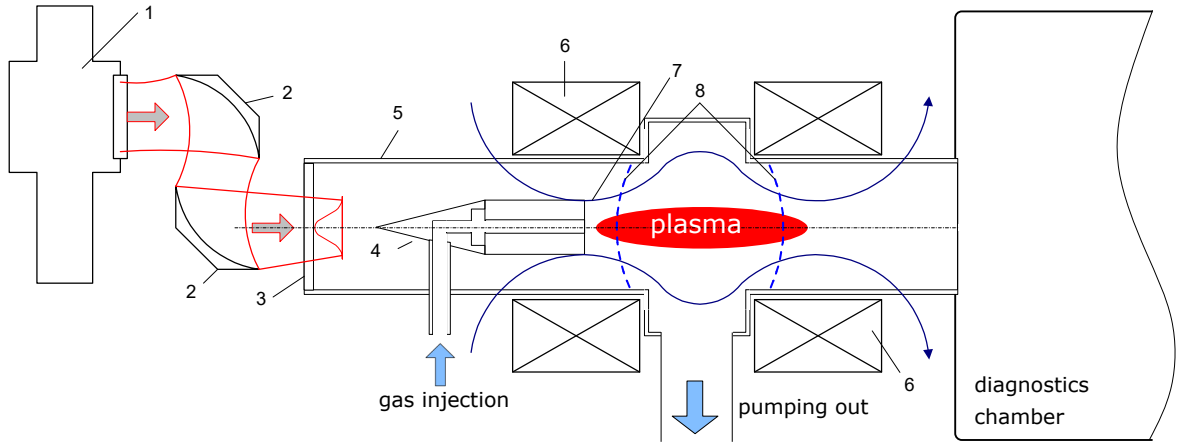}
	\caption{{The schematic view of the setup. Microwave emission of the gyrotron (1) travels via a quasi-optical transmission line (2) and is injected through the input window (3) and matching device (4) along the axial axis into the discharge chamber (5) (shown type 1). The matching device is located in the magnetic mirror point and is also used as a plasma electrode. Magnetic coils (6) form a simple mirror trap configuration (7) with the ECR zones (8) located between mirror points and the trap center. Blue arrows show the direction of working gas injection and its pumping out. All the diagnostics is placed inside the diagnostics chamber as well as outside from the vacuum volume through viewing ports in it.} \label{fig:setup}}
\end{figure*}

The experiments were conducted with the plasma of ECR discharge sustained by gyrotron radiation (frequency 37.5\,GHz, power up to 80\,kW, pulse duration 1\,ms) in the axially symmetric open magnetic trap \cite{Shalashov_2017_PhP}.
{The schematic view of the setup is shown in Fig.~\ref{fig:setup}.}
Microwave radiation of gyrotron is launched into the discharge chamber along the trap axis {through the plasma matching device}. The radiation intensity in the focal plane is about 10\,kW/cm$^{2}$ and the average power density is 100\,W/cm$^{3}$.
We used two types of discharge chamber: (1) a cylindrical tube with inner diameter of 38\,mm which is widened in the central part by the tube with inner diameter of 72\,mm and of 50\,mm length and (2) a straight cylindrical tube with inner diameter of 38\,mm. {The second type of the discharge chamber allows us to change the relative position of magnetic coils which is not possible in the first case.}

\begin{figure}[tbh]
	\centering
	\includegraphics[width=85mm]{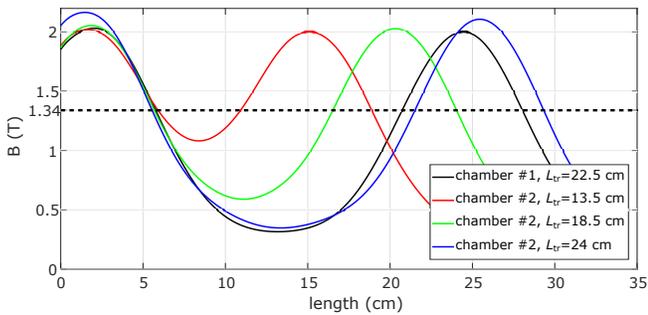}
	\caption{{Magnetic field distribution along the axis of the device for all configurations used in the experiments. The value $L_\mathrm{tr}$ shows the effective length of the magnetic trap. Horizontal dashed line denotes the resonant value of magnetic field at the ECR zone for the 37.5\,GHz gyrotron radiation. Data provided for the same voltage at the magnetic system charging device 3\,kV  and at the peak value of magnetic system current (3\,ms offset). Origin of horizontal axis corresponds to the position of the plasma matching device.}\label{fig:mf}}
\end{figure}

The discharge chamber is placed in the mirror magnetic trap, produced by pulsed coils with a maximum magnetic field strength of 4.3\,T, the pulse duration is about 7.3\,ms. The length of the magnetic trap is 22.5\,cm {with the mirror ratio $R=B_\mathrm{max}/B_\mathrm{min}\approx 5$} for the first type of discharge chamber and for the second type  discharge chamber the trap length is varied from 13.5\,cm to 24\,cm with a corresponding change of the mirror ratio. {The magnetic field distribution along the axis of the device for all configurations used in the experiments is shown in Fig.~\ref{fig:mf}. The value of the magnetic field strength in the specific magnetic configuration at the specific moment of time is defined by the voltage at the magnetic system charging device (from 1.5 to 4.5\,kV) and relative delay from the beginning of the magnetic field pulse. Varying these two parameters we may change absolute value of magnetic field and its time derivative.}

\begin{figure}[tbh]
	\centering
	\includegraphics[width=70mm]{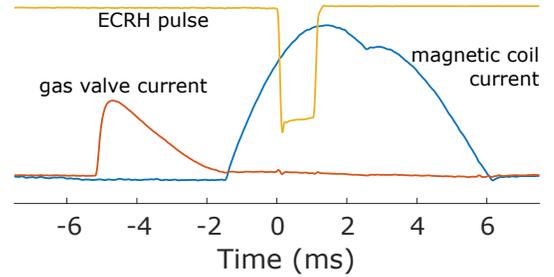}
	\caption{{Setup synchronization scheme. Time base corresponds to the beginning of the ECR heating pulse of 1 ms duration. The acquisition time for all the diagnostics used is set to 5\,ms.} \label{fig:sync}}
\end{figure}

Plasma is created and supported under ECR conditions at the fundamental cyclotron harmonic corresponded to the magnetic field strength 1.34\,T. The resonance surface is situated between the magnetic mirror  and the center of the discharge chamber {as shown in Fig.~\ref{fig:mf}. Neutral gas (nitrogen or argon) is injected into the discharge chamber by the pulsed valve. The delays between the gas valve open, ECR heating pulse and the magnetic field turn on are selected for the most favorable breakdown conditions. Setup synchronization scheme is shown in Fig.~\ref{fig:sync}.} Ambient pressure of a neutral gas is about $10^{-6}$\,Torr, however it increases up to $10^{-4}-10^{-3}$\,Torr during ECR discharge.

\begin{figure*}[tbp]
	\centering
	\includegraphics[width=140mm]{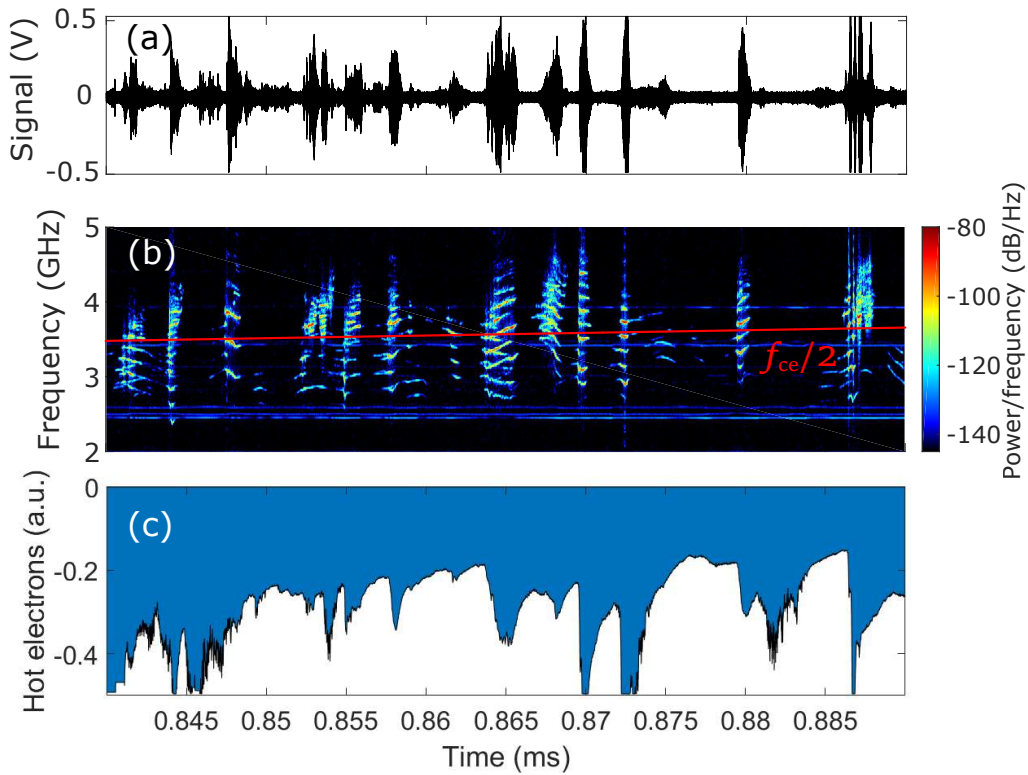}
	\caption{(a) Waveform and (b) its dynamic spectrum of plasma microwave emission during the stationary ECR discharge, (c) the signal from the hot electron detector {(negative polarity)}. The value of $f_{\mathrm{ce}}/2$ at the trap center is shown by the red solid curve on the dynamic spectrum. The time reference starts at the moment the gyrotron is turned on, the gyrotron radiation turns off after 1 ms. The working gas is nitrogen.\label{fig:whist1}}
\end{figure*}

In the experiments, we study the dynamic spectrum and the intensity of enhanced electromagnetic radiation from the plasma with the use of a broadband horn antenna with a uniform bandwidth in the range from 2 to 20\,GHz and the high-performance oscilloscope Keysight DSA-Z\,594A (analog bandwidth 59\,GHz, sampling rate 160\,GSample/s). In the scope of the described experiment, the antenna bandwidth covers frequencies up to the second harmonic of electron gyrofrequency in the trap center. This method allows to analyze both the wave amplitude and its phase. 
The dynamic spectra are calculated from the recorded data by short-time Fourier transform windowed with Hamming window.
Simultaneously,  we measure precipitations of energetic electrons (10-180\,keV)  from the trap ends using a p-i-n diode detector with time resolution about 1\,ns.

\section{Experimental results}

In the present paper we consider the developed discharge stage, which lasts from the ECR breakdown until the end of the microwave heating pulse with a duration of 1\,ms. At that stage the two-component electron population is formed containing the cold dense electrons with Maxwellian distribution ($N_\mathrm{c} \sim 10^{13}\,\mathrm{cm^{-3}}$, $T_\mathrm{c} \sim 300$\,eV) and less dense component of hot electrons with anisotropic distribution function ($N_\mathrm{h} \sim 10^{11}\,\mathrm{cm^{-3}}$, mean energy $E_\mathrm{h} \sim 10$\,keV, although energy tail is stretched up to 400\,keV, see, e.g., Ref.~\onlinecite{Izotov_RSI_2012}). 

During the developed discharge stage we registered microwave emission
simultaneously in two frequency bands: (1) at frequencies about $f_\mathrm{ce}/2$ and (2) at frequencies between $f_\mathrm{ce}$ and $2f_\mathrm{ce}$, where $f_\mathrm{ce}$ is the electron cyclotron frequency at the trap center. Here we discuss only the first type of microwave emission, which has much higher spectral power density than the second one.
{Initial experimental studies of this type of instability in a dense plasma of ECR discharge were done in Ref.~\onlinecite{avod2005} using microwave detectors, while the study of the fine structure of the spectrum was not yet possible.}

\begin{figure*}[th]
	\centering
	\includegraphics[width=150mm]{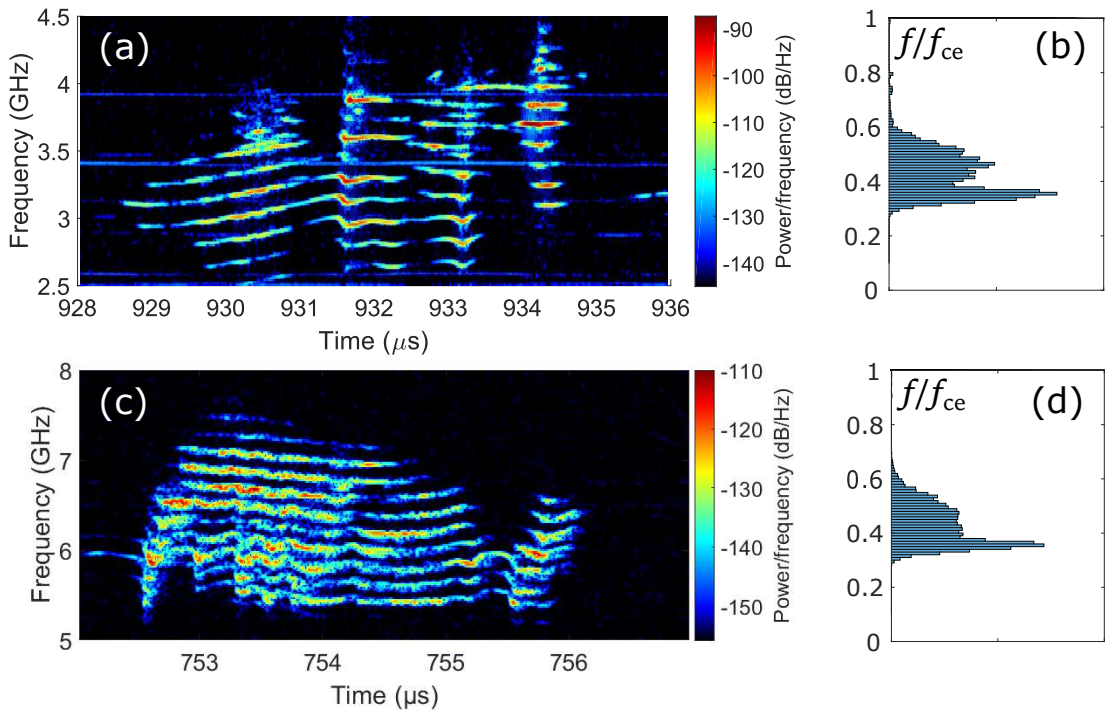}
	\caption{An example of the dynamic spectrum of plasma microwave emission during a single event of the instability development in (a) nitrogen and (c) argon plasmas. The time reference starts at the moment the gyrotron is turned on, the gyrotron radiation turns off after 1 ms. Panels (b) and (d) show the distribution of instability events over frequency for nitrogen (b) and  argon (d) discharges. Experiments were conducted with the discharge chamber type (1) -- a cylindrical tube which is widened in the central part.
	\label{fig:fine_hist}}
\end{figure*}

An example of the instability development is shown in Fig.~\ref{fig:whist1}. The recorded emission has the form of quasi-periodic bursts with a duration of about 1-5\,$\mu$s. The emission frequency follows $f_\mathrm{ce}/2$ while the external magnetic field is slowly increasing in time. The distinctive feature of this instability is the zebra-like pattern in the dynamic spectrum of plasma emission. Note the presence of the selected frequency lines, ten or more, in the spectrum, which are arranged equidistantly with respect to each other. These frequencies are slightly evolving in time while the distance between them remains constant.  
Every radiation pulse is correlated with precipitations of energetic electrons from the trap ends measured with a dedicated p-i-n diode.

The detailed view of the emission spectrum for a single onset of instability is shown in Fig.~\ref{fig:fine_hist} (left panels) for discharges in nitrogen and argon. It is seen that instability development lasts about 5\,$\mu$s and spectrum contains at least ten equidistantly spaced stripes for both cases. 

To measure the emission bandwidth we analyzed 54 experimental shots in nitrogen plasma and 131 in argon plasma. The spectrograms were filtered using a noise reduction algorithm. As a result, we obtained a set of spectrograms where only intense emissions are present. Next, all data points in a time-frequency domain were normalized over the electron cyclotron frequency in the trap center, which is evolving in time. Finally, we concatenated normalized spectrograms for a certain type of gas and retrieved a distribution of the events over frequency in the electron cyclotron frequency domain. These results are shown in Fig.~\ref{fig:fine_hist} (right panels). For both cases, the distribution  is two-humped with a more pronounced low-frequency maximum near $0.3 f_\mathrm{ce}$ and a wider second maximum near $0.5 f_\mathrm{ce}$.

Above mentioned results were obtained in a discharge chamber type (1) -- a cylindrical tube which is widened in the central part, where the length of the trap and its mirror ratio are fixed. 
To understand the formation mechanism of the zebra-like patterns, we studied the microwave emission of plasma created in the discharge chamber in the form of a straight cylinder -- type (2). We studied the instabilities during the ECR discharge in two gases (nitrogen and argon) and for a set of three different lengths of the magnetic trap $L$: 13.5\,cm, 18.5\,cm, 24\,cm. With the increase of distance between the magnetic coils the mirror ratio $R$ is also increased. The mean distance between spectral components for different experimental parameters is shown in Table~\ref{tab:freq}. For some regimes, we have not detected the excitation of waves at frequencies below the electron gyrofrequency. This could be related to the efficiency of the ECR heating at different conditions.

\begin{table}[tbh]
	\caption{\label{tab:freq} The distance between spectral components for different experimental parameters.}
	\centering
	\begin{tabular}{lccc}
		\hline \hline 
		$L$ & $B_\mathrm{max}/B_\mathrm{min}$ & N & Ar \\ 
		\hline \hline 
		13.5\,cm & 2 & not detected & not detected \\ 
		\hline 
		18.5\,cm & 3.5 & $272\pm17$\,MHz & $322\pm19$\,MHz \\ 
		\hline 
		24\,cm & 6 & not detected & $182\pm12$\,MHz \\ 
		\hline \hline 
	\end{tabular} 
\end{table}

\section{Discussion}

First of all, we note that the electromagnetic eigenmodes of the metal vacuum chamber can not be attributed to the observed spectra since this is not compatible with the equidistance of the bands in the spectrum and synchronous temporal variation of spectral lines.

The emission of dense plasma at frequencies below the electron cyclotron frequency during the  ECR heating stage is most naturally related to the whistler wave instability \cite{avod2005}. At large densities of the background plasma, the cyclotron instabilities of the extraordinary waves are suppressed because their dispersive properties are strongly modified by the background plasma.

\begin{figure}[tb]
	\centering
	\includegraphics[width=75mm]{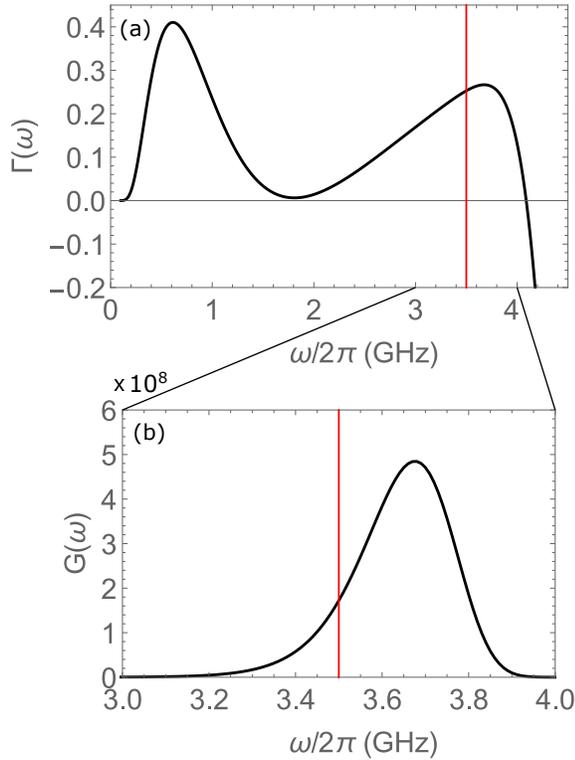}
	\caption{{(a) One-pass amplification gain $\Gamma(\omega)$ and (b) total amplification factor $G(\omega)=\exp[q\,\Gamma(\omega)]$ corresponding to $q=75$ passes trough the trap of a linearly unstable whistler wave caused by a butterfly-like electron distribution function in the presence of bulk background plasma.  Numerical simulation is done as described in Appendix for the following parameters: $n_\mathrm{c}=1.7\times10^{13}$\,cm$^{-3}$ (corresponds to the critical density for the gyrotron radiation at 37.5 GHz), $n_\mathrm{h}/n_\mathrm{c}=0.01$, $T_\perp/T_\parallel=5$, $T_\mathrm{c}/T_\parallel=0.025$, $T_\parallel=10$\,keV, $f_\mathrm{ce0}=7$\,GHz, $L=20$\,cm, $R=7$. Red vertical lines denote the half-gyrofrequency $f_\mathrm{ce0}/2$ at the trap center.
	} \label{fig:inc}}
\end{figure}

{

In conditions of the reported experiment, the whistler wave instability may be driven by the so-called ``butterfly-like'' electron velocity distribution which is formed during the off-center ECR heating due to consequent drift of fast electrons along the magnetic field lines towards the trap center (bounce oscillations) \cite{Shalashov_2017_PhP}. Unstable waves are traveling along the magnetic field and reflecting presumably from the metal surfaces at the trap ends. Since its one-pass amplification is likely to be relatively small, the waves need many passes over the instability region. In reported experiments, the estimated $q$-factor of the vacuum chamber is in the range of 50--100. Figure~\ref{fig:inc} shows an example of the one-pass amplification  gain and a total amplification factor corresponding to $q=75$ passes trough the trap volume corresponding to parameters close experimental conditions for Figs.~\ref{fig:whist1} and \ref{fig:fine_hist}(a). Details of these calculations are explained in the Appendix.  The one-pass gain has two broad peaks, one of which is in the  frequency range  from 2 to 4\,GHz which is detected in Fig.~\ref{fig:whist1}. Although small one-pass amplification realizes in a fairly  broad frequency range, the essential multi-pass gain is possible in a more narrow domain. In Fig.~\ref{fig:inc}(b), the total amplification factor has the maximum near the half of the electron gyrofrequency in the trap center and the peak width at half-maximum is about 200\,MHz, which corresponds to the spectral width of an individual line in the frequency spectrum in Fig.~\ref{fig:whist1}. 

Based on these observations we may assume that the formation of a frequency comb near the fundamental frequency is most likely a result of a low-frequency modulation of a medium where excited whistler waves could propagate.
}



For the majority of experimental shots the value of the electron cyclotron frequency in the trap center $f_\mathrm{ce}$ is in the range of 7--10\,GHz.
{For estimates we will use $f_\mathrm{ce}=7$\,GHz, cold plasma density $n_\mathrm{c}=1.7\times10^{13}$\,cm$^{-3}$ and $T_\mathrm{c}=250$\,eV.}
Correspondingly, the frequency of the ion cyclotron resonance is $f_\mathrm{ci}\approx 0.3$\,MHz for a nitrogen plasma and  $f_\mathrm{ci}\approx 0.1$\,MHz for an argon  plasma. The typical distance between spectral stripes  is $\Delta f\approx300$\,MHz, see Table \ref{tab:freq}. This, evidently, implies a very high ion cyclotron harmonics number $\Delta f/f_\mathrm{ci}\sim 10^3$, which ensures that ion cyclotron oscillations are hardly  related to the observed phenomenon.

\def\vv{\upsilon}
The other low frequency process is the rotation of the plasma column due to the presence of the electric field which is transverse to the magnetic field of the trap. We estimate the frequency  of $\mathbf{E}\times\mathbf{B}$-rotation as 
\begin{equation*}
f_{\mathrm{rot}}\approx\frac{\vv_\mathrm{E\times B}}{a}\approx\frac1{2\pi f_\mathrm{ce}}\frac{e\Delta\phi}{m_\mathrm{e} a^2}\lesssim 1\,\text{MHz}<\Delta f,
\end{equation*}
where $\vv_\mathrm{E\times B}=cE/B$ is the drift velocity, $E\approx\Delta\phi/a$ is the characteristic value of the radial electric field strength, $e\Delta\phi\approx 100$\,eV is the plasma potential drop across plasma column and $a=1$\,cm is the plasma radius. 

The excitation of whistler waves leads to the falling of the resonant electrons into the loss-cone. Then the abrupt ejection of electrons from the trap ends is observed. The electron precipitation causes the formation of {particle flows inside the trap which could excite ion-acoustic or magnetosonic} waves. 
Magnetosonic waves propagate with the Alfven velocity $\vv_A\sim {10^7}$\,cm/s. For the same perpendicular size of plasma column $a$ we obtain
\begin{equation*}
\label{eq:alfv}
f_\mathrm{A}\approx \frac{\vv_\mathrm{A}}{a}=\frac{B}{a\sqrt{4\pi n_\mathrm{i} m_\mathrm{i}}}\approx {(20-35)}\,\mathrm{MHz}<\Delta f,
\end{equation*}
the lower value of $f_\mathrm{A}$ corresponds to the argon plasma. 
{From these estimations, we may conclude that the magnetosonic waves, as well as the plasma rotation, could not lead to the observed frequency modulation of whistler waves.}

The frequency of ion-acoustic waves is limited by the ion plasma frequency:
{
\begin{equation*}
\label{eq:fpi}
f_{\mathrm{s}}\lesssim f_\mathrm{pi}=\frac{1}{2\pi}\quad \sqrt{\frac{4\pi  Z^2 e^2n_\mathrm{i}}{m_\mathrm{i}}},
\end{equation*}
where $Ze$ is the average ion charge,  $n_\mathrm{i} = n_\mathrm{e}/Z$ is the ion density, $m_\mathrm{i}$ is the ion mass.
In a high-power ECR discharge, the formation of multiply charged ions is highly probable  \cite{Golubev_RSI_2000, Skalyga_RSI_2016}. Basing on previous experience for the same setup, the average ion charge number in our experiments is estimated as  $Z=2-4$. Corresponding values of the ion plasma frequency $f_\mathrm{pi}$ are listed in Table~\ref{tab:zz}. One can see that these values  cover the frequency range of interest (200--300\,MHz) provided that $Z\gtrsim2$.  
Thus, we conclude that the observed modulation of whistler emission  could be eventually related to the scattering of waves by the low-frequency ion-acoustic waves. 

\begin{table}[tbh]
	\caption{\label{tab:zz} The ion plasma frequency for different ion charges.}
	\centering
	\begin{tabular}{lccc}
		\hline \hline 
		$Z$ & N & Ar \\ 
		\hline \hline 
		1 & 231\,MHz & 137\,MHz \\ 
		\hline 
		2 & 327\,MHz & 194\,MHz \\ 
		\hline 
		3 & 400\,MHz & 237\,MHz \\ 
		\hline 
		4 & 463\,MHz & 274\,MHz \\ 
		\hline \hline 
	\end{tabular} 
\end{table}

With this assumption, the distance between spectral components would depend on the plasma density, the average ion charge and the ion mass. In our experiment, the plasma density is defined mostly by the fixed frequency of ECR heating radiation: namely, the electron density $Z n_\mathrm{i}$ stays close to the cut-off density independent of all other conditions.  
On the other hand, we could expect the distances between the frequency stripes scales as $\sqrt{Z/m_\mathrm{i}}$ for different gas masses. Unfortunately, we have not enough experimental data to check this scaling law since the average ion charge is not measured during the present campaign. 

}

A more precise study of the zebra-like structures in the whistler frequency range to support this idea is subject to our future efforts.


\section{Summary}

The fine structure of dynamic spectra of electron cyclotron emission in whistler wave frequency range is investigated with high temporal resolution. The analysis of the experimental data suggests that formation of zebra-like patterns in the spectra could be related to the scattering of whistler waves by the low-frequency {ion-acoustic}  waves which could be excited during the abrupt ejection of electrons into the loss cone caused by the development of the electron cyclotron instability. 
{This assumption relatively well describes the equidistant frequency comb in the spectra. We suggest that the temporal variations of the spectrum are most probably  related to slow variations of cold plasma parameters, but this is still an open question to be analyzed in a future work.}

This research may be of interest in the context of a laboratory modeling of non-stationary processes of wave-particle interactions in space plasma, since there are a lot of open questions about the origin of some types of emissions in space cyclotron masers, especially mechanisms of fine spectral structure \cite{Bingham_2013,Melrose_2017}.

\begin{acknowledgments}
The work was supported by the Grants Council of the President of the Russian Federation (grant no.~MK-2593.2019.2). The authors thank anonymous referees for their efforts in improving the manuscript.
\end{acknowledgments}

{
\appendix*
\section{Calculation of the whistler wave instability gain}

The linear growth-rate $\gamma\equiv\mathrm{Im}\,\omega$ of the whistler wave instability is calculated for the non-relativistic regime in 
Ref.~\onlinecite{trakhtengerts_rycroft_2008}:
\begin{multline}\label{eq:inc}
\gamma=\frac{2\pi^3e^2}{m_\mathrm{e}c\: \partial(N_{||}\omega)/\partial\omega}\int \left(\pderv{F}{\upsilon_{||}}+\frac{\omega_{\mathrm{ce}}}{k_{||}\upsilon_\perp}\pderv{F}{\upsilon_\perp}\right) \times \\
\times\:\delta(\omega-k_{||}\upsilon_{||}-\omega_{\mathrm{ce}})\:\upsilon_\perp^3\mathrm{d}\upsilon_\perp\mathrm{d}\upsilon_{||}, 
\end{multline}
where $F(\upsilon_{||},\upsilon_\perp)$ is the electron distribution function  over parallel and perpendicular velocities to the external magnetic field normalized over particle density, $\delta$-function defines the cyclotron resonance  condition, $\omega=2\pi f$ is a cycle frequency, $k_{\parallel}=N_{||}\omega/c$ is a wave-vector component along the  magnetic field, $N_{||}$ is a corresponding refractive index, and $c$ is the speed of light. Formally this equation is obtained for the homogeneous plasma, but it may be applied in a weakly inhomogeneous limit assuming that the plasma parameters, the distribution function and the magnetic field are slowly varying functions along the trap axis $z$ compared to the wavelength of the unstable waves. Then, we may define the local spatial amplification coefficient $\eta\equiv 2\mathrm{Im}\,k_{||}$ and the amplification gain $\Gamma$ for a single passage of the wave from one end of the magnetic trap to another as 
\begin{equation}\label{eq:gain_int}
\eta(\omega,z)=2\gamma/\upsilon_\mathrm{gr},\quad 
\Gamma(\omega) = \int\displaylimits_{-L/2}^{L/2}\eta(\omega,z)\mathrm{d}z,
\end{equation}
 where $\upsilon_\mathrm{gr}=\partial\omega/\partial k_{||}$ is a wave group velocity. When the wave travels $q$ times between the trap ends, its intensity is amplified by $\exp(q\Gamma$) times. Hereafter we will use approximations 
$N_{||}\approx\omega_\mathrm{pe}/\sqrt{\omega(\omega_\mathrm{ce}-\omega)}$ and
$\upsilon_\mathrm{gr}\approx 2c\,(1-\omega/\omega_\mathrm{ce})^{3/2}\sqrt{\omega\omega_\mathrm{ce}}/\omega_\mathrm{pe}$
valid for the whistler wave propagating strictly parallel to the magnetic field. 

Strong ECR plasma heating in  adiabatic magnetic traps results in formation of anisotropic distributions of accelerated fast electrons characterized by a predominance of transverse velocities (relative to the direction of the magnetic field) to the longitudinal ones. For weakly relativistic electrons this may by modeled by bi-Maxwellian distribution, 
\begin {equation}\label {eq_edf} 
F=A\exp\left(-\frac{m_{\mathrm{e}} \upsilon_{||}^2}{2T_{||}}-\frac{m_{\mathrm{e}} \upsilon_{\perp}^2}{2T_{\perp}}\right).
\end{equation}
where $A$ is normalizing factor.
The feature of our experiments is that ECR heating is shifted from the trap center towards the higher magnetic field. 
Let the ECR heating corresponds to magnetic field $B=B_{\mathrm{ECR}}$. Following Liouville's theorem, the distribution function at the magnetic field $B'<B$ may be found as $F'(\upsilon'_{||},\upsilon'_\perp)=F(\upsilon_{||},\upsilon_\perp)$ where $\upsilon_{||}^2={\upsilon'_{||}}^2+{\upsilon'_{\perp}}^2(1-B/B')$ and $\upsilon_{\perp}^2={\upsilon'_{\perp}}^2 B/B'$. For the bi-Maxwellian  distribution (\ref{eq_edf}) one obtains
\begin {equation}\label {eq_edf1} 
F'=A\exp\left(-\frac{m_{\mathrm{e}} {\upsilon'_{||}}^{2}}{2T_{||}}-\frac{m_{\mathrm{e}} {\upsilon'_{\perp}}^2}{2T_{\perp}'}\right)\!\widetilde\Theta\left(\frac{\upsilon'_{\perp}}{\upsilon'}\right),
\end{equation}
where $\widetilde\Theta$  is a step-like function that defines the empty region  $\upsilon'_\perp/\upsilon>\sqrt{B'/B_{\mathrm{ECR}}}$ from which particles can not reach the ECR zone (the anti-loss-cone), and new effective transverse temperature is found from
$$\frac1{T'_{\perp}}=\frac1{T_{||}}\left(1-\frac {B_{\mathrm{ECR}}}{B'}\right)+\frac1{T_{\perp}}\frac {B_{\mathrm{ECR}}}{B'}.$$
Note that the effective temperature can be even negative for small enough magnetic field: $T'_\perp<0$ when $B'<(1-T_{||}/T_\perp)\:B_{\mathrm{ECR}}$. So, a strong enough anisotropy of the fast electrons in the heating zone results in butterfly-like distribution in  central parts of the trap that possess $\partial F/\partial \upsilon_\perp>0$ everywhere except the boundary of the anti-loss-cone \cite{Shalashov_2017_PhP}. It is natural to normalize the distribution function \eqref{eq_edf1} over the density of fast electrons measured at the trap center where magnetic field strength takes the minimal value $B'=B_0$:
\begin{equation*}
\int\limits_{\upsilon'_\perp/\upsilon<\sqrt{B_0/B_{\mathrm{ECR}}}} 2\pi \upsilon'_\perp F'\,  \mathrm{d}\upsilon'_\perp\mathrm{d}\upsilon'_{||}=n_\mathrm{h}.
\end{equation*}
Than $$A=\frac{n_\mathrm{h}}{(2\pi)^{3/2} T_{||}^{1/2} T'_\perp}\left(1-\sqrt{\frac{T_\perp}{T_{||}}\left(1-\frac{B_0}{B_{\mathrm{ECR}}}\right)}\right)^{-1}.$$
Note that $A>0$ even when $T'_\perp<0$.
The background plasma is modeled with the isotropic Maxwellian distribution function with density $n_\mathrm{c}$ and temperature $T_\mathrm{c}$, 
\begin {equation}\label {eq_edf2} 
F_\mathrm{M}=\frac{n_\mathrm{c}}{(2\pi T_{\mathrm{c}})^{3/2}}\,\exp\left(-\frac{m_{\mathrm{e}} \upsilon^2}{2T_{\mathrm{c}}}\right),
\end{equation}
which does not vary along the magnetic field. 

For the needs of the present work, the on-axis magnetic field may be approximated as
\begin{equation}\label{eqCf}
B(z)=\frac{B_0}{2}\left((1+R)+ (1-R)\cos\frac{2\pi z}L\right),
\end{equation}
where $B_0$ is the magnetic field strength at the trap center, $L$ is the end-to-end trap length, and $R=B_{\max}/B_{0}$ is the mirror ratio. For further convenience we will use dimensionless function $b(z)=B(z)/B_\mathrm{ECR}$ and constants $R_\mathrm{ECR} = B_\mathrm{ECR}/B_0$, $\omega_\mathrm{ECR}=2\pi f_\mathrm{ECR}$.

Substituting sum of distribution functions (\ref{eq_edf1}) and (\ref{eq_edf2}) into (\ref{eq:gain_int}), we obtain the expression for the amplification coefficient in an explicit analytic form:
\begin{equation}\label{eq:eta}
\eta(\omega,z) = \frac{n_\mathrm{h}}{n_\mathrm{c}}\left(\eta_1+\eta_2\right)+\eta_\mathrm{M}
\end{equation}
with
\begin{multline*}
\eta_1 = K\,\frac{\alpha +\left(1-\alpha\right)\omega_\mathrm{ECR}/\omega}{\left(1+\alpha (b-1)\right)^2} \times \\
\qquad\times b^2 x\left[\left(1+(\beta-1)x^2\right)\exp\left(-\beta x^2\right)-\exp\left(-x^2\right)\right],\\
\eta_2 =- K \,\frac{\omega_\mathrm{ECR}}{\omega}  \beta b^2 x^3\exp\left(-\beta x^2\right),\\
\eta_\mathrm{M} =- \sqrt{\pi}k_{||}\sqrt{\frac{ T_{||}}{T_\mathrm{c}}}x\exp\left(-\frac{T_{||}}{T_\mathrm{c}}x^2\right),\\
K =\sqrt{\pi} k_{||}  \frac{\alpha+\left(1-\alpha\right)R_\mathrm{ECR}}{1-\sqrt{\alpha\left(1-1/R_\mathrm{ECR}\right)}},\\
x=\frac{\omega_\mathrm{ce}-\omega}{k_{||}\sqrt{2 T_{||}/m_\mathrm{e}}},\;\;
\alpha=\frac{T_{\perp}}{T_{||}},\;\;
\beta = \frac{T_{||}}{T_{\perp}\left(1-b\right)}.
\end{multline*}
Here $\eta_1$ describes the deposition of the body of the fast electron distribution function (\ref{eq_edf1}); this term may be either positive (amplification) or negative (absorption). The last two terms, $\eta_2$ and $\eta_\mathrm{M}$, are always negative. Term $\eta_2$ describes additional damping due to abrupt jump of the distribution function (\ref{eq_edf1}) at the boundary of the anti-loss-cone  $\upsilon'_\perp/\upsilon=\sqrt{b}$; due to some physical reasons we take into account only $\partial F/\partial \upsilon_\perp$ derivative here. Term $\eta_\mathrm{M}$ describes the cyclotron damping in the background plasma. 

To get the whistler wave amplification (or absorption) for the one pass through  the magnetic trap, expression (\ref{eq:eta}) is integrated numerically over $z$ assuming that the magnetic field, i.e.\ parameters $b$, $x$ and $\beta$, vary according to (\ref{eqCf}). An example of such calculations for parameters relevant to our experiment is presented in Fig.~\ref{fig:inc}.

}

\bibliography{viktorov_php_2019}

\end{document}